# CO-EVOLUTION OF ATMOSPHERES, LIFE, AND CLIMATE

**John L. Grenfell[1,2*], Heike Rauer[1,2], Franck Selsis[3], Lisa Kaltenegger[4], Charles Beichman[5], William Danchi[6], Carlos Eiroa[7], Malcolm Fridlund[8], Thomas Henning[9], Tom Herbst[9], Helmut Lammer[10], Alain Léger[11], René Liseau[12], Jonathan Lunine[13], Francesco Paresce[14], Alan Penny[15], Andreas Quirrenbach[16], Huub Röttgering[17], Jean Schneider[18], Daphne Stam[19], Giovanna Tinetti[20], Glenn J. White[15,21]**

[1]DLR – German Aerospace Center, Institute of Planetary Research, Berlin, Germany
[2](*Now at) Technical University (TU) Berlin, Germany
[3]University of Bordeaux 1, Bordeaux, France
[4]Harvard Smithsonian Center for Astrophysics, Cambridge, MA, USA
[5]NASA ExoPlanet Science Institute, California Inst. Of Technology/JPL, USA
[6]Goddard Space Flight Center, Greenbelt, MD, USA
[7]Universidad Autonoma de Madrid, Madrid, Spain
[8]RSSD, ESA, ESTEC, Noordwijk, The Netherlands
[9]Max-Planck Institut für Astronomie, Heidelberg, Germany
[10]Space Research Institute, Austrian Academy of Sciences, Graz, Austria.
[11]Universite Paris-Sud, Orsay, France
[12]Dept. of Radio and Space Science, Chalmers University of Technology, Onsala, Sweden
[13]Lunar and Planetary Laboratory, University of Arizona, USA
[14]INAF,Via del Parco Mellini 84, Rome, Italy
[15]Space Science & Technology Department, CCLRC Rutherford Appleton Laboratory, Oxfordshire, UK
[16]Landessternwarte, Heidelberg, Germany
[17]Leiden Observatory, Leiden, The Netherlands
[18]Observatoire de Paris-Meudon, LUTH, Meudon, France
[19]SRON, Netherlands Institute for Space Research, Utrecht, The Netherlands
[20]Department of Physics and Astronomy, *University College London, London, UK*
[21]The Open University, Milton Keynes, UK

Corresponding Author:
John L. Grenfell
E-mail: lee.Grenfell@dlr.de
DLR – German Aerospace Centre,
Institute of Planetary Research, Berlin, Germany

## ABSTRACT

After Earth's origin, our host star, the Sun, was shining 20 to 25 percent less brightly than today. Without greenhouse-like conditions to warm the atmosphere, our early planet would have been an ice ball and life may never have evolved. But life did evolve, which indicates that greenhouse gases must have been present on early Earth to warm the planet. Evidence from the geologic record indicates an abundance of the greenhouse gas $CO_2$. $CH_4$ was probably present as well, and in this regard methanogenic bacteria, which belong to a diverse group of anaerobic procaryotes that ferment $CO_2$ plus $H_2$ to $CH_4$, may have contributed to modification of the early atmosphere. Molecular oxygen was not present, as is indicated by the study of rocks from that era, which contain iron carbonate rather than iron oxide. Multicellular organisms originated as cells within *colonies* that became increasingly specialized. The development of *photosynthesis* allowed the Sun's energy to be harvested directly by life forms. The resultant oxygen accumulated in the atmosphere and formed the *ozone* layer in the upper atmosphere. Aided by the absorption of harmful UV radiation in the *ozone layer*, life colonized Earth's surface. Our own planet is a very good example of how life forms modified the atmosphere over the planets' life time. We show that these facts have to be taken into account when we discover and characterize atmospheres of Earth-like exoplanets. If life has originated and evolved on a planet, then it should be expected that a strong co-evolution occurred between life and the atmosphere, the result of which is the planets' climate.

Keywords: Early Earth, bio-marker, atmospheres, climate, exoplanets.

## 1. HOW LIFE HAS AFFECTED EARTH'S ATMOSPHERE

To estimate the occurrence of terrestrial exoplanets and maximize the chance of finding them, it is crucial to understand the formation of planetary systems in general and that of terrestrial planets in particular. A reliable formation theory should not only explain the Solar System, with small terrestrial planets within a few AU and gas giants further out, but also the newly discovered planetary systems with close-in giant planets. Regarding the presently known exoplanets, it should be stressed that our current knowledge is strongly dependent upon the sensitivity limits of the current detection techniques (mainly the radial velocity method). With time and improved detection methods, the diversity of planets and orbits in extrasolar planetary systems will definitely increase and help to constrain the formation theory further.

Nitrogen on Earth was outgassed during the first hundred million years. Therefore, the atmospheric pressure was at least 0.8 bar in Earth's prebiotic atmosphere. If climate regulation via the carbonate-silicate cycle is assumed (Walker et al. 1977), then the level of $CO_2$ is determined to a first approximation by the solar luminosity.

- Due to the faint luminosity of the early Sun (e.g., Gough 1981; Baraffe et al. 1998), about 200-300 mbar of $CO_2$ would have been necessary to ensure a mean surface temperature above 273 K (Kasting 1993).

Molecular hydrogen was likely the third main component of Earth's prebiotic atmosphere; In an enhanced presence of $CO_2$ and if large amounts of hydrogen were absent in the upper atmosphere, the exosphere may have been relatively cool, which could have resulted in low escape rates of atomic hydrogen (Tian et al. 2005). Hydrogen that was released by volcanoes but not efficiently lost to space must have accumulated to levels of the order of 200 mb (Tian et al. 2005). As water was present on Earth before 4.4 Gyr, water vapor was also an important constituent of the lower atmosphere. Other atmospheric species that may have been present at this time include: CO and sulphur-bearing species like $H_2S$ released by volcanoes and possibly methane produced abiotically in hydrothermal vents.

### *1. 1. A rise in methane?*

Among the most primitive Archaea found in the tree of life, as shown in Fig. 1, are the methanogens, some some of which are autotrophic (consuming $CO_2$ and $H_2$) and others heterotrophic (consuming organic molecules). Methane is a trace gas in the present Earth atmosphere (about 2 ppm), and its origin is biological except for a small fraction produced in hydrothermal systems.

- Methanogens existed almost certainly during the Archaean and the Neoproterozoic, when the atmosphere was still anoxic (before 2.3 Gyr).

If we assume a biogenic release equal to the present day, the level of methane would have reached 100 - 1000 PAL (Present Atmospheric Level) in the absence of atmospheric $O_2$ (Pavlov et al. 2000). As today's methanogens can only grow in very limited environments where $O_2$ is absent and $H_2$ or organics are present, the production of methane by the biosphere was probably much higher in the early anaerobic environment. Thus, very high levels of methane can be inferred, which lasted for more than 1 Gyr, between the emergence of methanogens (probably earlier than 3.4 Gyr) and the rise of $O_2$ (2.3 Gyr).

- Such high levels of $CH_4$ would have had a strong impact on climate and geochemical cycles. Methane is a very efficient greenhouse gas, and levels higher than 100 PAL would potentially have been enough to warm the surface above 0°C (Pavlov et al. 2000), which implies that climate is no longer regulated by the carbonate-silicate cycle. The level of $CO_2$ could potentially have been extremely low when methane became the main greenhouse gas, and this seems to be confirmed by the studies of paleosoils from the late Archaean and Neoproterozoic, in which no trace of carbonates were found (Rye et al. 1996; Hessler et al. 2004). Note that this does not take the effect of hazes into account. In the early atmosphere, the haze expected from $CH_4$ photolysis was probably weak if the $CH_4/CO_2$ ratio did not exceed unity. If the ratio had exceeded unity, an organic haze would have formed (Pavlov et al. 2003) such that it would have potentially cooled the atmosphere and regulated the greenhouse effect. Another important consequence of high levels of $CH_4$ is the transport of hydrogen, which would have dissociated from methane by high XUV radiation of the young Sun to the upper atmosphere and led to a much higher escape rate of hydrogen to space.

Thus, it might be possible that biological methanogens contributed to the oxidation of the atmosphere and lithosphere and enhanced the loss of H, making possible, later, for the rise of oxygen (Carling et al. 2001).

### *1. 2. The build-up of $O_2$*

Geological records have revealed the chemical action of free oxygen after about 2.3 Gyr ago (Bekker et al. 2004), except for some deposits from the deep ocean that remained anoxic for a few hundred million years or more (Rouxel et al. 2005).

- It seems that the build-up of atmospheric $O_2$ occurred at least 400 Myr after the emergence of oxygen-producing bacteria capable of oxygenic photosynthesis (Fig. 2).

Indeed, 2.7 Gyr old molecular fossils are interpreted as the remains of primitive cyanobacteria and eukaryotes, which are producers and consumers of $O_2$, respectively (Brocks et al. 1999). Several reasons could explain this delay. First, the budget reaction of oxygenic photosynthesis also works in the reverse direction, since respiration and oxidation of organic sediments consume oxygen.

- Thus, the build up of atmospheric $O_2$ requires the burial of the organics produced by photosynthesis.

This occurs today at a rate of 589 Tg $O_2$/year (Catling and Claire 2005), which means that the net release of the present $O_2$ atmospheric content ($10^{21}$g $O_2$) takes about 2 Myr (this is about 1000 times slower than the production of $O_2$ by photosynthesis). This rate is balanced by the oxidation of rocks, old sediments, and volcanic gases. The oxidation sinks for $O_2$ may have been much more efficient on early Earth, partly due to the presence of large amounts of reduced iron in the ocean and the crust (Walker 1977).

- In the absence of efficient organic burial, $O_2$ would not build up.

Some tectonic processes may have favored the burial of reduced carbon and allowed the rise of $O_2$ by about 2 Gyr ago (Des Marais et al. 1993). Another hypothesis has already been mentioned and is linked with the slow oxidation of Earth through the escape of hydrogen to space: in other words,

- the Earth had to be depleted of a lot of hydrogen before an oxygen atmosphere could occur.

There might also be a climatic reason for the delay between the emergence of $O_2$-producers and the rise of $O_2$. $CH_4$ has a very short photochemical lifetime in an $O_2$-rich atmosphere, which means that a consequence of a build up of $O_2$ is a decrease of the $CH_4$ atmospheric abundance and, thus, a fall of the mean surface temperature that could lead to a global freezing event.

Therefore, in a biosphere where $CH_4$ and $O_2$ producers both exist, the solar luminosity might be a strong constraint on the timing of the oxygenation (Selsis 2002).

- All these constraints on the build up of an $O_2$-rich atmosphere are extremely important for astrobiological considerations, as some authors argue that complex multicellular life can only develop in an oxic environment (e.g., Catling et al. 2005).

Geological records have provided us with only qualitative information about the presence or absence of oxygen in the atmosphere. After the rise of $O_2$ and until the end of the Precambrian (about 550 Myr ago), it can only be inferred that the level of $O_2$ was above about 1% PAL (0.2 % in abundance). During the Phanerozoic (from - 550 Myr to now), models based on the chemical and isotopic composition of sedimentary rocks allow us to trace back the evolution of the level of $O_2$ and show that it has varied roughly between 0.7 and 1.8 PAL (Berner et al. 2003).

- The highest oxygen level (nearly twice that of today, which implies an atmospheric pressure 15% higher) is found to have occurred during the Permo-Carboniferous.

The principal cause of this enhanced level was the rise of large vascular land plants and the consequent increased global burial of organic matter. Higher levels of $O_2$ are consistent with the presence of Permo-Carboniferous giant insects.

### *1. 3. The rise of the ozone layer*

Ozone ($O_3$) is produced by only one reaction: $O + O_2 + M \rightarrow O_3 + M$, where M is any compound (this is called a 3-body or association reaction) and atomic oxygen comes from the photolysis of $O_2$. On the other hand, $O_3$ is destroyed by photolysis and many trace compounds in the atmosphere ($HO_X$, $NO_X$, $ClO_X$, …).

- Therefore, the amount of $O_3$ in the atmosphere depends on the level of $O_2$ but also on the abundance of these trace gases on which we do not have enough information to infer the level of $O_3$ in the Precambrian.

The variation of $O_2$ alone during the last 550 Myr would not have changed significantly the level of $O_3$, as ozone has only a weak dependency on the $O_2$ level (Léger et al. 1993; Segura et al. 2003).

The abundance of $O_3$ was more certainly affected by changes in the trace gases content. We do not know for sure whether, between the rise of $O_2$ and the beginning of the Phanerozoic, $O_3$ provided a UV shield for land life, but it can be inferred that it was present when the first lichens colonized the lands during the Ordovician (500 - 425 Myr ago).

## 2. STELLAR RADIATION AND CLIMATE: IMPLICATIONS FOR ATMOSPHERIC BIOMARKERS

It is well known from satellite observations and the output of coupled chemistry-climate models on Earth that important physical factors like atmospheric temperature and solar radiation affect atmospheric biomarker molecule concentrations over diurnal and seasonal timescales.

- Since it is expected that a terrestrial exoplanet-finding mission will discover terrestrial exoplanets for various stellar spectral-types and stellar evolution phases, how different radiation inputs and climate feedbacks can affect biomarker signals must be studied.

As a first step, changes to biomarkers in Earth's atmosphere in response to variations related to orbital parameters, climate feedbacks, and the radiation and particle environment of the Sun/stars can be studied with the use of theoretical climate models (Fig. 3).

*2.1. Ozone ($O_3$)*

Tropospheric $O_3$ makes up about 10% of the overhead column content in the terrestrial atmosphere. In the lower troposphere (0 - 8 km), $O_3$ is formed when volatile organic compounds such as $CH_4$ are oxidized in the presence of $NO_x$ (see overview in Crutzen 1988), and it is destroyed by surface deposition. In the mid to upper troposphere (8 – 18 km), stratospheric-tropospheric exchange affects $O_3$ concentrations. Here, stratospheric air may fold down into the troposphere and release $O_3$-rich air. Stratospheric $O_3$ makes up about 90% of the $O_3$ layer that forms when $O_2$ is photolyzed (< 242 nm) into O, which then reacts with $O_2$ to form $O_3$.

- Why does an $O_3$ layer form at a specific altitude?

In the lower atmosphere, there is little $O_3$ because UV is weak; in the upper atmosphere, there is little $O_3$ because there is little $O_2$. Fig. 4 shows that the trade-off between UV and $O_2$ availability with altitude leads to a sharply defined $O_3$ maximum in the tropical stratosphere at 10mb (30km), which travels each year across the equator toward the summer hemisphere where photolysis rates are higher.

In the lower stratosphere (18 - 25 km, 50 - 100 mb), chemical timescales that affect $O_3$ are long compared with dynamical timescales. Thus, $O_3$ behaves as a quasi-inert tracer of dynamical motion in that it is quasi-horizontal along isentropic (constant entropy) surfaces with typical stratospheric horizontal and vertical diffusion coefficients of 1000 and 0.015 $m^2/s$ respectively (e.g., Waugh et al. 1997).

In the upper stratosphere (35 – 45 km, 5 – 1 mb), $O_3$ is controlled mainly by chemistry. The main source is $O_2 + O + M \rightarrow O_3 + M$ whereas important sinks are $O + O_3 \rightarrow 2O_2$ and $O_3 + h\nu(< 320\ nm) \rightarrow O + O_2$. These reactions were first discussed in depth by Chapman (1930) and are usually referred to as Chapman chemistry.

- On the Earth, $O_3$ is formed in the tropics via Chapman chemistry and then transported slowly (over several months) to mid and high latitudes.

Chapman chemistry alone would produce more than twice as much $O_3$ as is measured. To explain the discrepancy, catalytic cycles that destroy $O_3$ were proposed:

$$X + O_3 \rightarrow XO + O_2$$

$$XO + O \rightarrow X + O_2$$

---

Overall: $O_3 + O \rightarrow 2O_2$

where X can be OH (Bates and Nicolet 1950), NO (Crutzen 1970), and Cl (Molina and Rowland 1974). The cycles are referred to as $HO_x$, $NO_x$ and $ClO_x$ cycles. Stratospheric $HO_x$ comes, e.g., from $H_2O$ degradation, $NO_x$ from $N_2O$ photolysis, and $ClO_x$ from manmade chlorofluorocarbons (CFCs). $HO_x$ cycles are generally important in the tropical lower stratosphere, $NO_x$ in the middle, and $ClO_x$ in the upper stratosphere (e.g., Lary 1997 and references therein).

- Stratospheric $O_3$ features a natural cycle with around 20 % more column $O_3$ in spring than in autumn.

The spring maximum is associated with less photolytic $O_3$ loss over the preceding winter. At polar latitudes, the natural cycle is opposed by the $O_3$ hole that forms via chemical processes in spring with typically 30 – 60 % loss in the column. Over short timescales (days to weeks) the $O_3$ column in mid to high latitudes is affected by atmospheric dynamics (Staehelin 1998).

In the mesosphere (50 - 85 km), intense UV leads to a diurnal cycle with daytime conversion of $O_3$ to $O^3P$. Observations by Zommerfelds et al. (1989) suggest a factor of 3 variation in $O_3$ at 65 km and a factor of 6 at 74 km. $HO_x$ cycles are an important sink for mesospheric $O_3$ (Summers et al. 1996).

### *2. 2. Water ($H_2O$)*

On Earth, the vast majority of atmospheric $H_2O$ resides in the troposphere.

- The region around the tropopause acts as a cold trap that prevents most $H_2O$ from reaching the stratosphere.

Fig. 5 shows the $H_2O$ concentration as does Fig. 4 for $O_3$ as a function of zonal-mean height. Sherwood (2001) reviewed processes that regulate the entry of $H_2O$ from the troposphere into the stratosphere. Strong convective cloud events in the tropics achieve this if they penetrate the lower stratosphere. Methane oxidation is an important source of stratospheric $H_2O$ and is initiated by reaction with hydroxyl (OH): $CH_4 + OH \rightarrow CH_3 + H_2O$.

Mesospheric $H_2O$ features an annual cycle with amplitude of ~ 20 % and maximum in summer linked with stronger upward transport of damp air (Jackson et al. 1998).

### *2.3. Methane ($CH_4$)*

On Earth, between 80 – 90 % of $CH_4$ resides in the troposphere where it is well-mixed. Fig. 6 shows the $CH_4$ concentration as a function of zonal-mean height. During summer, when $CH_4$ oxidation is faster, the total column values are 5 – 10 % lower than winter. $CH_4$ is removed in the troposphere and stratosphere by reaction with the hydroxyl (OH) radical. This is a major source of water in the stratosphere.

- About one third of $CH_4$, i.e., $225 \times 10^{12}$g C/ year (225 Teragrammes, Tg) of $CH_4$ emissions arise naturally at the surface via, e.g., geologic activity (Etiope and Klusman 2002) or methanogenic bacteria under anaerobic conditions (wetlands and oceans) (Intergovernmental Panel on Climate Change IPCC, 2001: Krüger et al. 2001).

- The remaining two-thirds arise from human activity (industry and agriculture).

- Stratospheric $CH_4$ is rather inert, with a typical lifetime of 5-10 years, so it acts as a tracer of dynamical air motions.

This can be seen in Fig. 6, which shows contours curving upward (rising air) in the tropics and downward (sinking air) toward the pole. This is the so-called equator-to-pole circulation of the Earth, which will be discussed later. Mesospheric $CH_4$ is low, due to stratospheric oxidation.

### *2.4. Nitrous Oxide ($N_2O$)*

Like $CH_4$, $N_2O$ is most abundant in the troposphere. It is extremely inert with a lifetime of around 150 years and, hence, is a good indicator of dynamical motions.

- 7 - 14 $\times$ $10^{12}$g N/year $N_2O$ are emitted into the atmosphere (Oonk and Kroeze 1998) mainly via denitrifying bacteria.

In the stratosphere, about 95 % $N_2O$ is removed via photolysis in the stratosphere. The remaining 5 % reacts with $O^1D$ to form NO.

- A breakdown of $N_2O$ contributes to $O_3$ destruction via $NO_x$ cycles.

Little $N_2O$ survives into the mesosphere, where concentrations are about 500 times less than they are in the stratosphere.

*2.5.Atmospheric dynamics and temperature*

On Earth, the total $O_3$ column is sensitive to atmospheric dynamics because a large amount of $O_3$ resides between 18 - 25 km, where chemical loss is slow. Other biomarkers occur mainly in the troposphere, where they are well-mixed, so their column values are less sensitive to dynamical influence.

- Enhanced solar heating in the tropics coupled with the Earth's rotation sets up an equator-to-pole flow called the world-, meridional-, or Brewer-Dobson (Brewer 1949) circulation (Holton 2004).

On Earth, in a 10-year cycle, air enters the lower stratosphere in the tropics, travels up into the summer mesosphere, then back across the equator into the winter hemisphere, moving poleward and downward. The circulation is stimulated by atmospheric waves (Haines 1991), e.g., gravity waves that can form when air flows over mountains or planetary waves which form via conservation of potential vortices when air moves in the north-south plane on a rotating globe.

Fig. 7 shows the $O_3$ concentration in parts per million by volume as a function of temperature. Up to 100 mb air cools due to adiabatic expansion. Above that height, radiative heating, mainly from $O_3$, leads to a temperature inversion and marks the start of the stratosphere. Between 1 - 0.1 mb, this heating subsides and cooling with height marks the start of the mesosphere.

## 3. IMPLICATIONS FOR TERRESTRIAL EXOPLANETS

Earth's climate is non-linear with complex feedbacks. It is challenging enough for current climate models to reproduce accurately, e.g., the $O_3$ response to climate change over the past twenty years (see the IPCC Third Assessment Report on Climate Change, 2001 for an overview). Several groups have investigated Earth's atmospheric evolution (Kasting and Catling 2003, Pavlov et al. 2000) to reduce atmospheres for which differing levels of $CH_4$ and $CO_2$ are assumed to the current day atmosphere. Spectral modes of these evolution scenarios show different detectable features over Earth's geological history in low resolution (Kaltenegger et al. 2008).

Ozone in the troposphere: sensitivity computational model runs for varying solar flux as well as $CH_4$, $O_2$, and $NO_x$ concentrations appropriate to the proterozoic period of the early Earth indicated that the tropospheric $O_3$ column could sometimes constitute appreciably more than today's 10 % of the total column (Grenfell et al. (2006)). The additional $O_3$ was produced via the photochemical smog mechanism, in which $CH_4$ oxidation is catalyzed by $NO_x$. The mechanism therefore favors terrestrial exoplanets with:

- weak stratospheric $O_3$, therefore high UV levels that reach the troposphere,

- abundant $CH_4$ from volcanoes, methanogens, or both,

and

- abundant $NO_x$ (from lightning and the action of cosmic rays on $N_2/O_2$ atmospheres).

Ozone photolysis effect: For Earth, the standard stratospheric Chapman chemistry analysis assumes both a rapid inter-conversion of O and $O_3$, and O in a steady-state so that rate of O production = rate of O loss, $j(O_2) = k\,(O)(O_3)$, $(O_3) = j(O_2)/(O)k$, where j and k are photolysis and reaction coefficients, respectively, with units of $s^{-1}$.

- For terrestrial exoplanets, these assumptions may not apply because the rate of O production and O loss depend, e.g., on temperature, UV radiation, etc.; therefore, one has to apply in first steps a simpler approach.

The reaction $O + O_3 \rightarrow 2O_2$ quickly slows at low temperatures; therefore, one can assume a cold case, in which this reaction can be neglected. In such a case, $O_3$ is destroyed directly by $h\nu < 320$ nm but is formed when $O_2$ dissociates at $h\nu < 242$ nm (UVC).

Therefore, under such conditions, an interesting quantity is the ratio, R = [UV-B (280 - 315nm) / UV-C (100-280nm)].

- If, on a terrestrial exoplanet, the atmospheric radiative transport differs potentially from Earth-like conditions, then regions that have small R values would experience weak $O_3$ loss (from direct $O_3$ photolysis in the UV-B) but strong $O_3$ formation (from $O_2$ photolysis hence $O_3$ formation in the UV-C).

The total incoming flux at the top of the atmosphere, R, is:

- R = 2.31 (Sun), R = 2.98 (K2V) and , R=0.92 (F2V) stars (Segura et al. 2003).

More studies are required to explore this theme further and establish where, or if, such effects can be valid.

Ozone Temperature Effect: That $O_3$ features a negative correlation with temperature has long been recognized (WMO 1998). This behavior arises because the $O_3$ sink, $O + O_3 \rightarrow 2O_2$, speeds up considerably as temperature increases. The temperature dependency can be seen from the rate constant of the reaction, k, where $k = 8.0 \times 10^{-12} \exp^{(-2060/T)}$ (DeMore et al. 2003). Substituting T=200, 250, 300 K implies $k = (2.7, 21.1 \text{ and } 83.4) \times 10^{-16}$ molecules$^{-1}$cm$^{3}$s$^{-1}$, respectively, i.e.,

- It can be expected that a large increase in the $O_3$ sink occurs as T increases.

This result is relevant to warm or hot exoplanets, e.g., close to the inner edge of the habitable zone (HZ) or with strong greenhouse warming, or both.

Ozone and NOx: Exoplanets in the HZ with M-dwarf hosts may have weak magnetospheres and, therefore, higher levels of cosmic ray bombardment. Griessmeier et al. (2005) suggested that the percent of cosmic rays that reach such a planet's surface to be a factor of 2 - 10 greater, compared with the Earth, assuming the two worlds have similar surface areas.

- Enhanced cosmic ray events imply increased atmospheric $NO_x$ loadings because the cosmic ray particles can lead to ionization of $N_2$ (Siskind et al. 1997).

- Such events on Earth typically lead to a 50 – 100 % increase in $NO_x$ in the upper stratosphere together with 10 – 30 % $O_3$ loss (Sinnhuber et al. 2003; Jackman et al. 2000; Quack et al. 2001).

The amount of total $O_3$ column loss depends on how easily $NO_x$ can propagate down into the mid-stratosphere, where $O_3$ is most abundant.

Water and $HO_x$: Kasting (1993) suggested that, near the inner edge of the HZ, all tropospheric $H_2O$ may eventually overcome the tropopause cold trap and flood into the stratosphere.

- In such a scenario, destruction via $HO_x$ cycles would then likely destroy most of the $O_3$ layer.

In general, this mechanism suggests that a temperature increase at the surface leads to faster evaporation and, hence, higher atmospheric humidity. The effect is difficult to quantify in climate models due to opposing mechanisms (e.g., more $H_2O$ leads to more clouds, which can produce a cooling effect) and interfering processes (e.g., stratospheric $H_2O$ also comes from $CH_4$ oxidation, the cold trap temperature is influenced by other factors) (IPCC 2001). One can anticipate that ocean-world (Leger et al. 2003) exoplanets are potentially most vulnerable to such a mechanism.

In addition to forming $NO_x$, cosmic rays may also lead to atmospheric $HO_x$ formation. Solomon et al. (1981) suggested the mechanism for this, which involves an $H_2O$ ion-cluster. Similar to the $NO_x$ effect discussed previously, the amount of total $O_3$ column loss depends on how easily $HO_x$ can propagate down into the mid-stratosphere, where $O_3$ is most abundant.

- $H_2O$ is particularly sensitive to Lyman-alpha (121.6 nm) radiation, whereby it rapidly photolyses to form $HO_x$. Chandra et al. (1997) suggested 30 – 40 % and 1 – 2 % change in terrestrial $H_2O$ at 80 km and 60 km, respectively, in response to Lyman-alpha changes.

Absolute Lyman-alpha fluxes are especially large for A-stars (and to a lesser extent, F-stars). They are about 30 times larger during the first 100 Myr after G-type stars have arrived at the zero-age-main-sequence (Ribas et al. 2005), so their exoplanets are the most likely candidates to lose their atmospheric $H_2O$ via this mechanism, assuming a low opacity atmosphere.

Ozone and Dynamics: Terrestrial exoplanets with stronger differential heating gradients compared to those of Earth will feature a stronger equator-to-pole circulation.

- On Earth, for example, slowing this circulation would imply a change in the amount of $O_3$ found in the tropics (where it is formed) relative to the amount of $O_3$ at higher latitudes, to where it is transported. Other effects like strong temperature gradients would potentially influence such effects (Spiegel et al. 2008)

Orographic features (mountains) also play an important role. On Earth, the northern hemisphere (NH) is much more mountainous than the southern hemisphere (SH). Surface airflow over mountains leads to waves being excited, which carry heat and momentum up into the stratosphere and stimulate the equator-to-pole circulation, with the result that the NH winter stratosphere is typically 10 –20 K warmer than the SH (World Meteorological Organization, Report 44, 1998). There is, however, no data for this effect over Earth's history, so the effect is difficult to quantify for different topologies. On a cloud-free planet with surface features like those of Earth, the diurnal flux variation in the visible caused by different surface features rotating in and out of view could be high, assuming hemispheric inhomogeneity (Ford 2001; Seager and Ford 2002). When the planet is only partially-illuminated, compared to a fully-illuminated case, a more concentrated signal from surface features could be detected as they rotate in and out of view.

Methane ($CH_4$): Methanogen sources have quite specific temperature (and UV) dependencies that vary with species and environment (e.g. soil type ) (*Nozhevnikova et al. 2003)*. On the early Earth, atmospheric sinks and geologic sources may have been much stronger than today (Kasting and Catling 2003); therefore,

- terrestrial exoplanets with fast tectonic activity and low $O_2$ may favor higher $CH_4$,

while,

- terrestrial exoplanets with some $O_2$ may favor fast $CH_4$ removal via OH.

Nitrous Oxide ($N_2O$): $N_2O$ is formed via denitrification in soil, which, like methanogenesis, is dependent on the temperature and the soil type (Li et al. 1992).

- $N_2O$ is destroyed mainly via UV radiation, which suggests that strong UV fluxes could lead to strong photolytic $N_2O$ loss.

It is probably not useful to apply derived physical $N_2O$ relationships to exoplanet conditions since responses (e.g. of denitrifying bacteria) can be species-specific and finely-tuned to Earth conditions.

Biomarker molecules in the Earth atmosphere respond to complex feedbacks, which sometimes produce widely differing results, e.g., in chemistry-climate model responses [IPCC 2001]. Nevertheless, it can be assume that the main influences are sufficiently well-known to justify a discussion of biomarkers on Earth-like exoplanets (see Kaltenegger et al. this volume).

- $O_3$ is clearly sensitive to the UV radiation in that it is both formed and destroyed in different regions of the UV radiation, which depends on the radiative transfer in the atmosphere.

- Furthermore, $O_3$ concentrations are inversely proportional to temperature due to an increase in the sink: $O + O_3 \rightarrow 2O_2$. This will be especially important for terrestrial exoplanets near the inner HZ or with strong greenhouse warming.

- $O_3$ is destroyed catalytically by $HO_x$, $ClO_x$ and $NO_x$ - the latter may be particularly important for terrestrial exoplanets orbiting the HZ of M- and some K dwarves.

Ultimately, more sensitivity computer simulations that use atmospheric models are required to separate out the influences of the various feedbacks. $H_2O$, $CH_4$, and $N_2O$ present additional measurement difficulties since they mainly exist in the troposphere below the cloud-base.

- Warming the troposphere leads to more $H_2O$ overcoming the cold trap at the tropopause and hence reaching the stratosphere. If this process proceeds too quickly, it will lead to removal of the $O_3$ layer via $HO_x$ destruction.

Estimating changes in $CH_4$ and $N_2O$ sources on terrestrial exoplanets is difficult since, on the Earth, they vary from organism to organism and are adapted to their particular biological niche.

- Regarding $H_2O$, $CH_4$, and $N_2O$ sinks, enhanced stellar fluxes are expected to lead to a faster direct photolytic sink, as well as faster removal via ($CH_4$+OH) and ($N_2O$+$O^1D$), since the radicals OH and $O^1D$ are formed photolytically.

Regarding seasonal and diurnal chemical cycles, these are most easily detected for $O_3$ (again, because other biomarkers occur mainly in the troposphere, where photolysis rates are low). On the Earth, $O_3$ column seasonal variations are in the region of about 20 % for the natural cycle at mid-latitudes and about 50 % for human-induced $O_3$-hole variations poleward of about $60^o$.

- Clearly, when observing terrestrial exoplanets, the strength of the biomarker seasonal signal depends on the viewing geometry. Note that first generation space missions will observe disk integrated spectra of exoplanets with no spatial resolution.

If we view only one hemisphere, then the seasonal signal will be strong compared to instances where both hemispheres are viewed simultaneously. On Earth, a small seasonal $O_3$ signal still persists, even if we average data over both hemispheres. Austin (2002) suggested a 3% cycle in mean observed column $O_3$ (averaged from 65° S to 65° N, i.e., excluding the effects of the manmade $O_3$ hole), with peak values in NH spring (which does not translate into a feature that is remotely detectable). This result is related to hemispheric asymmetry – the meridional circulation is stronger in the north due to more wave stimulation from mountain ranges.

However, it is entirely feasible that seasonal and daily biomarker amplitudes for many terrestrial exoplanets could be greater than those of Earth. Larger obliquity, for example, leads to stronger seasonal variations in stellar flux and temperature. Increases in ellipticity leads to seasons of unequal duration, as on Mars, and larger differences between closest and furthest approach also leads to large amplitudes in seasonal fluxes and temperatures.

Bertrand et al. (2002) and Williams et al. (2003) performed sensitivity runs for an Earth-like exoplanet with changed orbital parameters (e.g., increasing ellipticity), using a climate model. The latter study suggested an intensely strong seasonality, e.g., equatorial surface temperatures varying from zero in winter up to 80°C in summer that would potentially be a much stronger influence on chemical variation.

## CONCLUSIONS

Studying biomarkers helps us understand the potential chemistry on exoplanets, different chemical cycles as well as observed spectra. It is worth exploring the parameter space of biomarkers because we only see a very limited range here on Earth.

## ACKNOWLEDGEMENTS

The authors acknowledge the Helmholtz-Gemeinschaft as this research has been supported by the Helmholtz Association through the research alliance "Planetary Evolution and Life", the International Space Science Institute (ISSI; Bern, Switzerland) and the ISSI teams "Evolution of Habitable Planets" and "Evolution of Exoplanet Atmospheres and their Characterization".

**Figure 1**

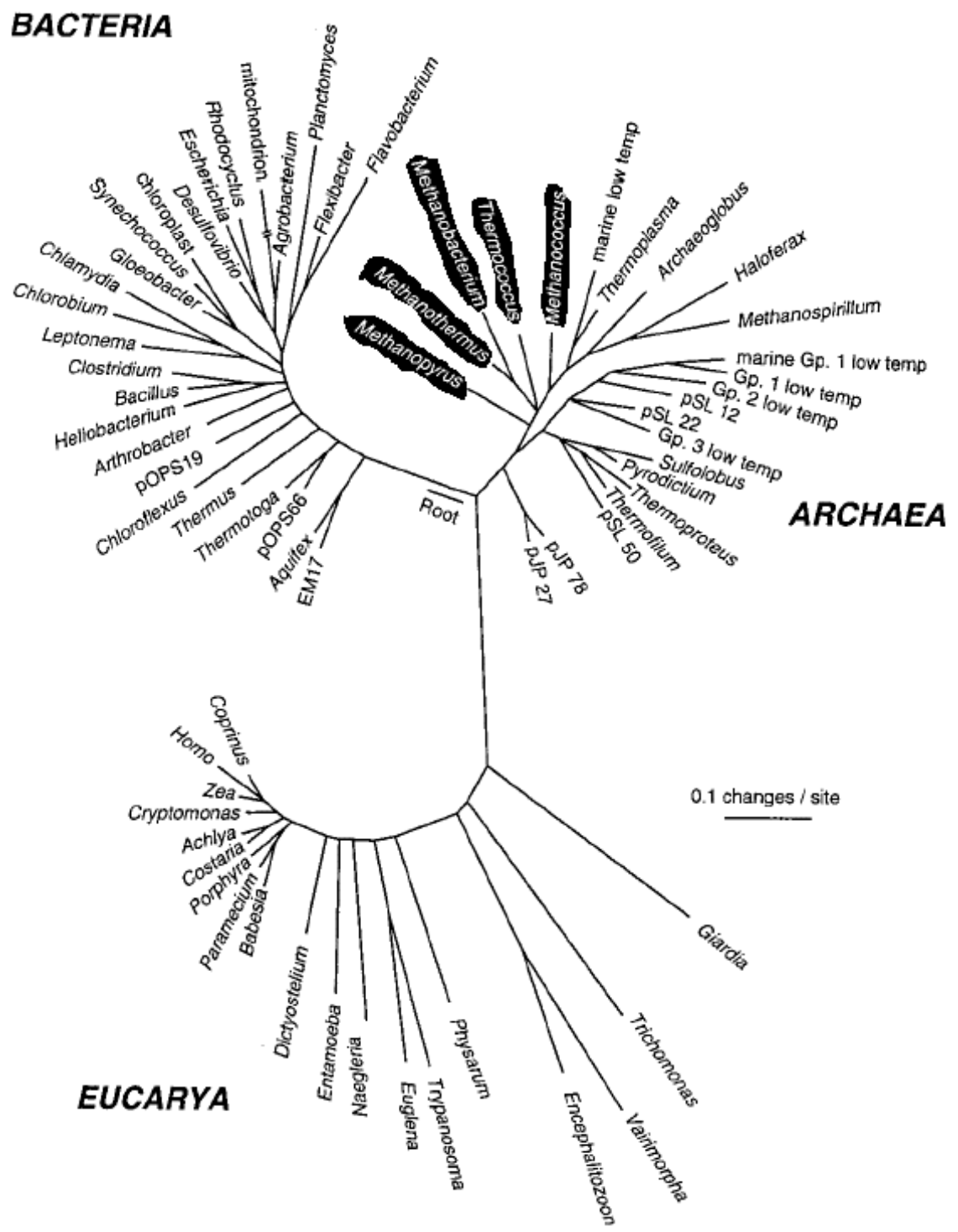

**Fig. 1:** Tree of life: in black are indicated some methanogens that are among the most primitive life-forms known (Courtesy of N. Pace).

**Figure 2**

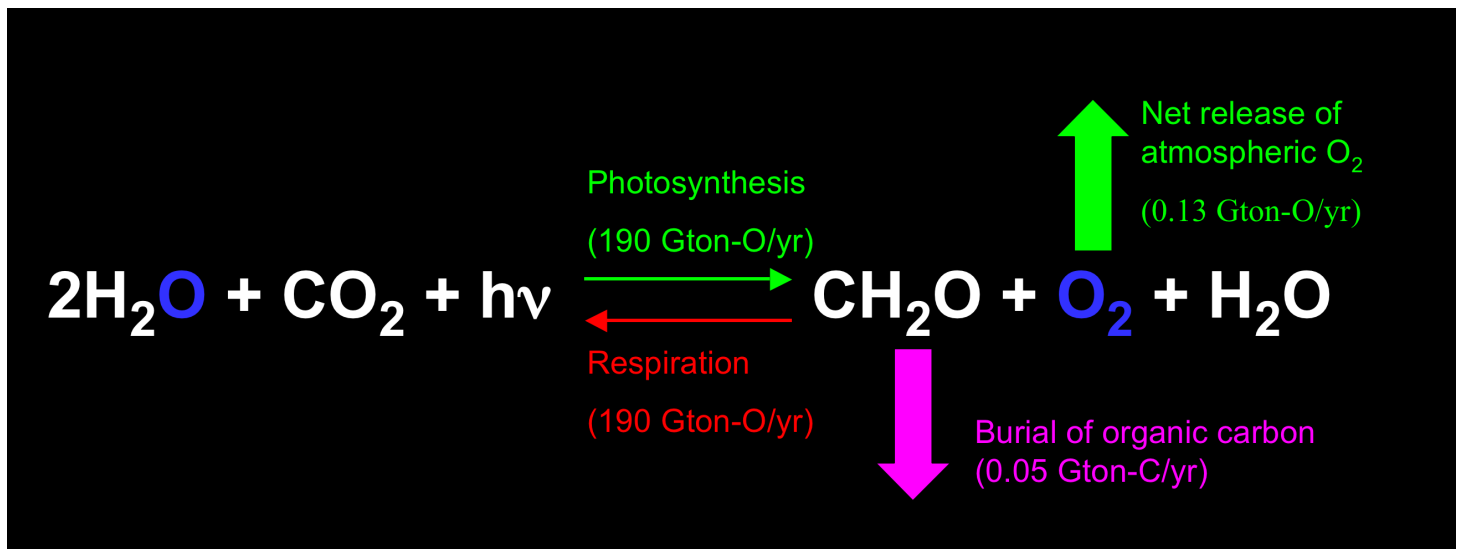

**Fig. 2:** Illustration of reactions that result in a release of $O_2$ in Earth's atmosphere.

**Figure 3**

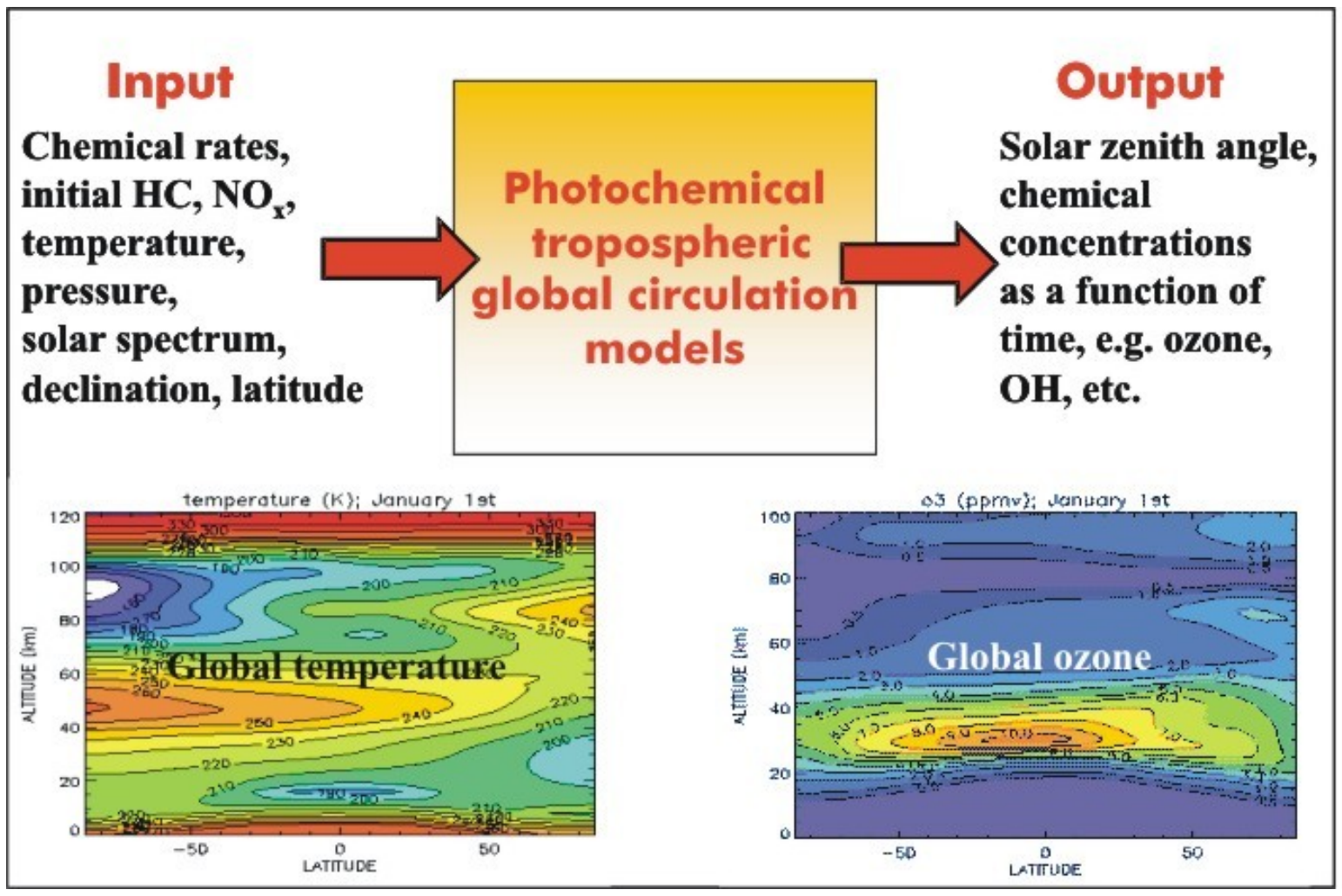

**Fig. 3:** Illustration of the application of coupled chemistry-climate models and sophisticated 3-D global circulation models (GCM), which are currently used in studies for comparative planetology on Solar System planetary bodies like Venus, Earth, Mars, and Saturn's satellite Titan. It is important to expand such studies to terrestrial exoplanets, because different radiation and particle fluxes related to different host star-types will effect the climate, chemistry, and expected biomarkers in the atmospheres (The results were taken from the SOCRATES model at the Atmospheric Chemistry Division at the National Center for Atmospheirc Research: http://acd.ecar.edu/models/SOCRATES).

**Figure 4**

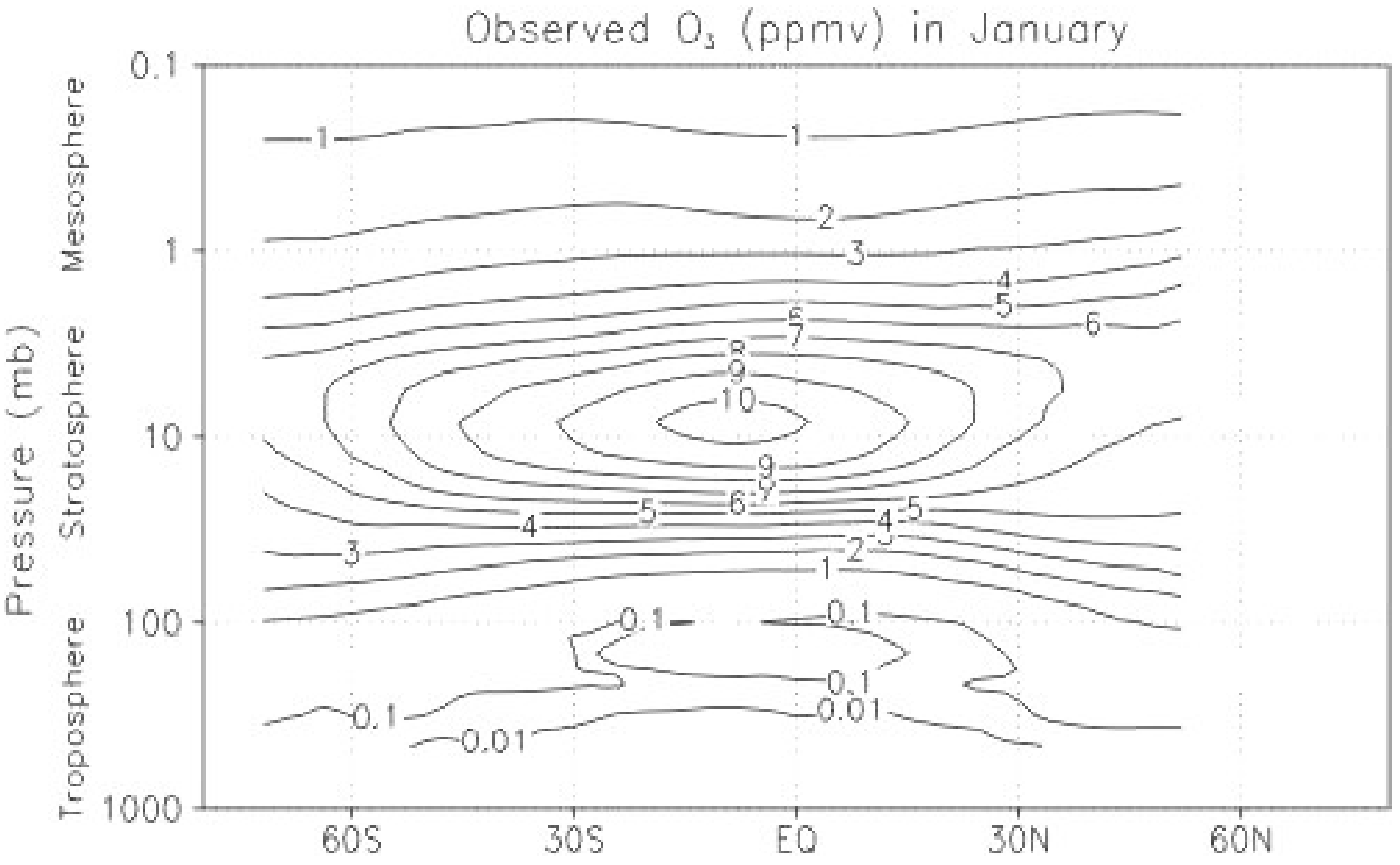

**Fig. 4:** $O_3$ concentration (parts per million by volume) shown as a function of zonal-mean altitude. Data represent the mean of 11 Januaries from 1992 to 2002, data level 2 version 19 measured by the Halogen Occultation Experiment (HALOE) aboard the Upper Atmosphere Research Satellite (UARS) (Russell et al. 1993).

**Figure 5**

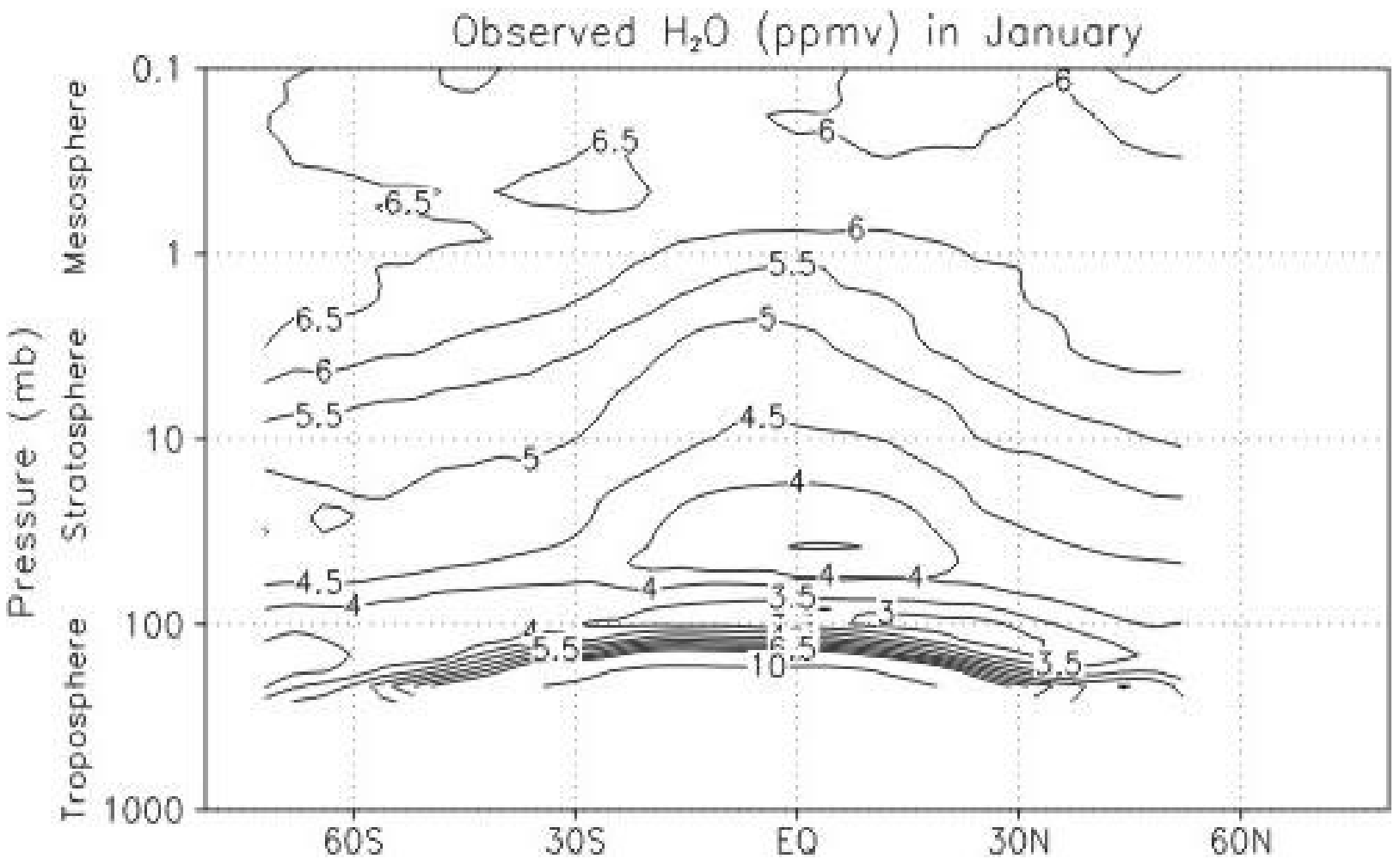

**Fig. 5:** Water concentration in parts per million by volume shown in Earth's atmosphere as a function of zonal-mean height.

**Figure 6**

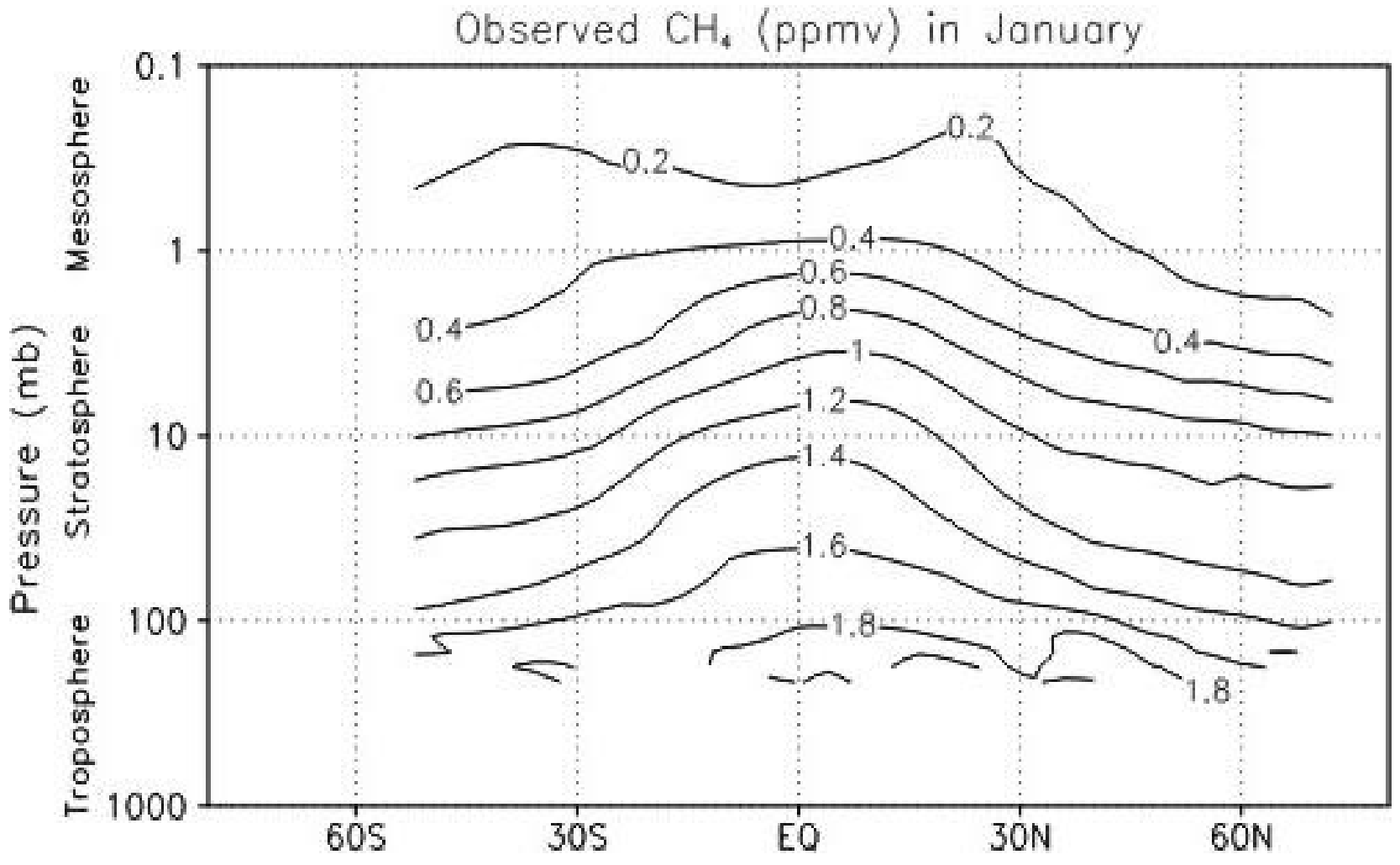

**Fig. 6:** Methane concentration in parts per million by volume shown in Earth's atmosphere as a function of zonal-mean height.

**Figure 7**

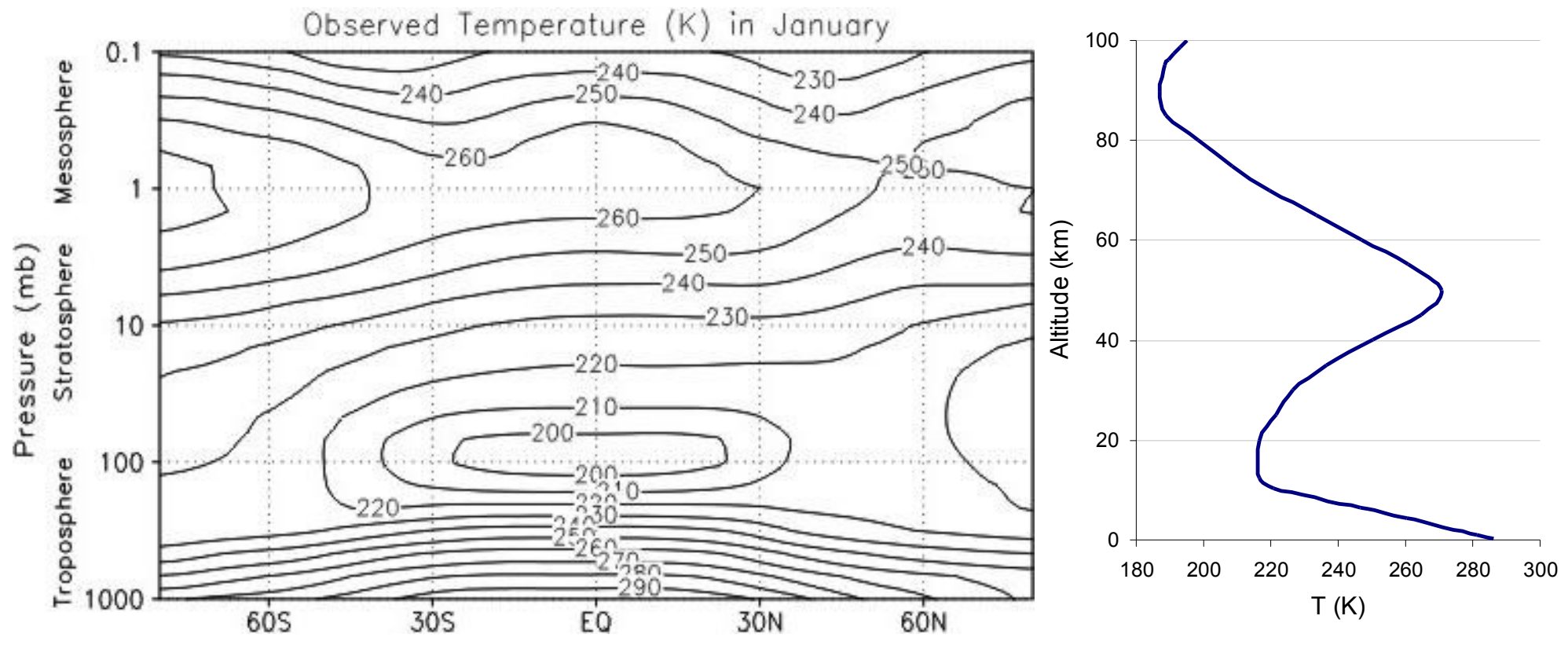

**Fig. 7:** $O_3$ concentration (parts per million by volume) shown as a function of temperature. Data shows the mean January value from 1991 to 2002, derived from various observational platforms and compiled as part of the "Stratospheric Processes and their Role in Climate" (SPARC) Program by Bill Randel at UCAR (Randel et al. 2004). (right) average temperature profile of US1978 standard atmosphere.